\documentclass[twocolumn,aps,prb,floatfix]{revtex4}
\usepackage[latin9]{inputenc}
\usepackage{graphicx}
\usepackage{esint}

\makeatletter
\@ifundefined{textcolor}{}
{%
 \definecolor{BLACK}{gray}{0}
 \definecolor{WHITE}{gray}{1}
 \definecolor{RED}{rgb}{1,0,0}
 \definecolor{GREEN}{rgb}{0,1,0}
 \definecolor{BLUE}{rgb}{0,0,1}
 \definecolor{CYAN}{cmyk}{1,0,0,0}
 \definecolor{MAGENTA}{cmyk}{0,1,0,0}
 \definecolor{YELLOW}{cmyk}{0,0,1,0}
}

\usepackage{bm}\usepackage{float}\usepackage{afterpage}

\makeatother

\begin{document}

\title{Dimensional evolution between one- and two-dimensional topological
phases}

\author{Huaiming Guo$^*$, Yang Lin}

\affiliation{Department of Physics, Beihang University, Beijing, 100191, China}

\author{Shun-Qing Shen}

\affiliation{Department of Physics, The University of Hong Kong, Pokfulam Road,
Hong Kong}
\begin{abstract}
Dimensional evolution between one- ($1D$) and two-dimensional ($2D$)
topological phases is investigated systematically. The crossover from
a $2D$ topological insulator to its $1D$ limit shows oscillating
behavior between a $1D$ ordinary insulator and a $1D$ topological
insulator. By constructing a $2D$ topological system from a $1D$
topological insulator, it is shown that there exist possibly weak topological phases
in $2D$ time-reversal invariant band insulators, one of which can be realized in anisotropic systems. The topological
invariant of the phase is $Z_{2}=0$. However the edge states may appear along specific
boundaries. It can be interpreted as arranged $1D$ topological phases,
and have symmetry-protecting nature as the corresponding $1D$ topological
phase. Robust edge states
can exist under specific conditions. These results provide further
understanding on $2D$ time-reversal invariant insulators, and
can be realized experimentally.
\end{abstract}


\maketitle

\section{Introduction}

Topological insulator (TI) is a novel quantum state of matter, which
is determined by the topological properties of its band structure.
It has generated great interests in the field of condensed matter
physics and material science due to its many exotic electromagnetic properties and possible potential
applications \cite{rev1,rev2,rev3,rev5}. Its discovery also deepens
understanding on the time-reversal invariant band insulators. In two
dimensions, ordinary insulator and quantum spin Hall insulator   are
characterized by a $Z_{2}$ invariant $\nu$: $\nu=0$ for a conventional
insulator and $\nu=1$ for a quantum spin Hall insulator \cite{kane1,kane2}.
In three dimensions ($3D$) time-reversal invariant band insulators
can be classified into $16$ topological classes distinguished by
four $Z_{2}$ topological invariants, and thus the ordinary insulator
is distinguished from 'weak' and 'strong' TIs \cite{kane3,kane4}.

The $3D$ TIs have been proposed and verified in many materials \cite{3d1,3d2,3d3,3d4,3d5}.
However the $2D$ TIs have only been realized in HgTe/CdTe and InAs/GaSb/AlSb
quantum well systems \cite{2d1,2d2,2d3}. Theoretical studies have
suggested that the $2D$ TI may be achieved in a thin film of $3D$
TI. In the thin film, the quantum tunneling between the two surfaces
generates a hybridized gap at the Dirac point. Depending on the thickness of the quantum wells,
the system oscillates between an ordinary insulator and quantum spin
Hall insulator \cite{film1,film2}. Typical TI materials, such as
$Bi_{2}Te_{3}$ and $Bi_{2}Se_{3}$ have a layered structure consisting of
weakly coupled quintuple layers, which makes it relatively easy to
grow high quality crystalline thin films using molecular beam epitaxy.
Until now the thin films of $3D$ TIs have been successfully fabricated
experimentally and the gap-opening has been observed \cite{grow1,grow2}.
This paves the way to realize the quantum spin Hall insulator from
the present various $3D$ TIs, which will greatly enlarge the family
of $2D$ TIs. Furthermore by introducing ferromagnetism in thin film,
the quantum anomalous Hall effect can be realized, which has been
experimentally confirmed in magnetic TIs of Cr-doped $(Bi,Sb)_{2}Te_{3}$
\cite{qah1,qah2}. Also it has been shown that the Chern number of
quantum anomalous Hall effect can be higher than one by tuning exchange
field or sample thickness\cite{qah3,qah4}. Compared to the $3D$
strong TIs, the weak TIs are related to the topological property of
the lower dimensions in a more direct way, since it can be interpreted
as layered $2D$ quantum spin Hall insulator. Though no weak TIs have
been reported experimentally, they are expected to have interesting
physical properties \cite{weak1,weak2,weak3,weak4,jiang}. Besides the studies
on the $2D$ limit of $3D$ TIs, recently a theoretical formalism
has been developed to show that a $3D$ TI can be designed artificially
via stacking $2D$ layers \cite{lim}. It provides controllable approach
to engineer 'homemade' TIs and overcomes the limitation imposed by
bulk crystal geometry.

The above studies show the connection between the $2D$ and $3D$
TIs. It is notable that recently there are increasing interests in
$1D$ topological phases \cite{1d1,guo1,guo2,guo3}. Specially they
have been studied experimentally using ultra-cold fermions trapped
in the optical superlattice and photons in photonic quasicrystals
\cite{1d2,1d3} and metamaterials \cite{1dsr}. With the developments
of these techniques, various $1D$ models with topological properties
may be realized \cite{1d4}. Also these techniques can be easily extended
to $2D$ or $3D$ cases. It is also desirable to study the $1D$ topological
phase in real materials. A natural thought is to narrow a $2D$ TI
and the narrow strip may be a $1D$ TI. Also a formalism on how to
construct $2D$ TIs from $1D$ ones is needed. The underlying question
is the connections between the $1D$ and $2D$ TIs.

In the paper, the question is studied systematically. It is found
that  the crossover from the $2D$ quantum spin Hall insulator to
its $1D$ limit shows oscillatory behavior between a $1D$ ordinary
insulator and $1D$ TI . Generally the $2D$ TI is a 'genuine' $2D$
phase and cannot be understood simply from the corresponding $1D$
phases. However by arranging $1D$ TIs, it is found that there exists
a $2D$ weak topological phase in anisotropic systems. In contrast to the quantum spin Hall
insulator: the $2D$ weak TI has topological invariant $Z_{2}=0$; the edge states are mid-gap ones and only appear along specific boundaries. These results provide further
understanding on $2D$ time-reversal invariant insulators, and
can be realized experimentally. The paper is organized
as the following: In Sec.\mbox{II},  the model Hamiltonian is introduced
to describe the $1D$ and $2D$ TIs; In Sec.\mbox{III}, an oscillatory
crossover from $2D$ to $1D$ topological phases is observed; In Sec.
\mbox{IV}, a $2D$ weak TI is identified by arranging $1D$ topological models; In Sec.
\mbox{V},  the physical properties of the $2D$ weak TI are studied;
Finally Sec. \mbox{VI} is the conclusion of the present paper.

\section{The $1D$ and $2D$ TI models}

\begin{figure}[htbp]
\centering \includegraphics[width=8.5cm]{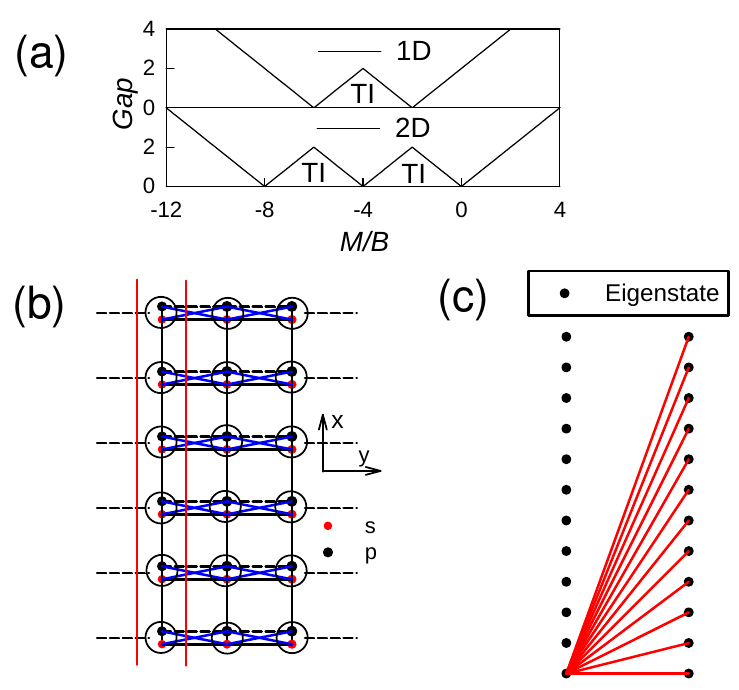} \caption{(a) The phase diagram of the Hamiltonian Eq.(\ref{eq1}) and Eq.(\ref{eq2}).
(b) The schematic diagram of the $2D$ model Eq.(\ref{eq1}), in which
the $A$-kind (blue solid lines) and $B$-kind (dashed and solid black
lines) hoppings along $y$-direction are shown. (c) The real hopping
between the NN chains are converted to the hopping between the eigenstates
of the NN chains. }

\label{fig1}
\end{figure}

The starting point of the present work is the 2D tight-binding model
for a quantum spin Hall insulator \cite{2d1,2d0},
\begin{eqnarray}
H_{2D} & = & \sum_{i}(M+4B)\Psi_{i}^{\dagger}I\otimes\sigma_{z}\Psi_{i}\label{eq1}\\
 & - & \sum_{i,\hat{x}}B\Psi_{i}^{\dagger}I\otimes\sigma_{z}\Psi_{i+\hat{x}}-\sum_{i,\hat{y}}B\Psi_{i}^{\dagger}I\otimes\sigma_{z}\Psi_{i+\hat{y}}\nonumber \\
 & - & \sum_{i,\hat{x}}sgn(\hat{x})iA\Psi_{i}^{\dagger}s_{z}\otimes\sigma_{x}\Psi_{i+\hat{x}}\nonumber \\
 & - & \sum_{i,\hat{y}}sgn(\hat{y})iA\Psi_{i}^{\dagger}I\otimes\sigma_{y}\Psi_{i+\hat{y}}\nonumber
\end{eqnarray}
where $I$ is the identity matrix and $\sigma_{j}$, $s_{j}(j=x,y,z)$ are the Pauli matrices representing
the orbit and spin, respectively; $\Psi_{i}=(s_{i\uparrow},p_{i\uparrow},s_{i\downarrow},p_{i,\downarrow})^{T}$
with $s_{i\uparrow(\downarrow)}$($p_{i\uparrow(\downarrow)}$) electron
annihilating operator at site $\mathbf{r}_{i}$. The first term
is the on-site potential, which has different signs for the $s-$
orbit and $p-$ orbit. The second and third terms are the hopping
amplitudes among the $s-$ orbits or $p-$ orbits, which are
differed by a sign. The third and fourth terms are the hopping amplitudes
between the $s-$ orbit and $p-$ orbit electrons, which is due to
the spin-orbit coupling. $A$ and $B$ are the hopping amplitudes
and in the following of the paper we take $B$ positive and set $A=1$
as a unit of the energy scale. The Hamiltonian is invariant under
time-reversal ${\cal T}=is_{y}\otimes I{\cal K}$. It belongs to the
AII class and its topological property is described by a $Z_{2}$
index \cite{class}. For $M>0,M<-8B$ it is a trivial insulator. For
$-8B<M<0$ it is a quantum spin Hall insulator with the topological
invariant $Z_{2}=1$. In the Hamiltonian, the subsystems of spin-up
and -down are decoupled.

Based on the spin-up subsystem and reducing one dimension (such as
the $y-$ dimension), a 1D spinless topological model is obtained
\cite{guo1,guo2,guo3},
\begin{eqnarray}
H_{1D} & = & \sum_{i}(M+4B)\Psi_{i\uparrow}^{\dagger}\sigma_{z}\Psi_{i\uparrow}-\sum_{i,\hat{x}}B\Psi_{i\uparrow}^{\dagger}\sigma_{z}\Psi_{i+\hat{x}\uparrow}\label{eq2}\\
 & - & \sum_{i,\hat{x}}sgn(\hat{x})iA\Psi_{i\uparrow}^{\dagger}\sigma_{x}\Psi_{i+\hat{x}\uparrow}.\nonumber
\end{eqnarray}
At half-filling, the system is a non-trivial insulator for $-6B<M<-2B$
and a trivial insulator for $M>-2B$ or $M<-6B$. In the momentum
space it becomes:$H_{1D}(k)=[M+4B-2Bcos(k)]\sigma_{z}+2Asin(k)\sigma_{x}$.
The Hamiltonian possesses a particle-hole symmetry $\sigma_{x}H_{1D}^{*}(k)\sigma_{x}=-H_{1D}(-k)$,
a pseudo-time-reversal symmetry $\sigma_{z}H_{1D}^{*}(k)\sigma_{z}=H_{1D}(-k)$
and a chiral symmetry $\sigma_{y}H_{1D}(k)\sigma_{y}=-H_{1D}(k)$.
It belongs to the BDI class and its topological invariant is a winding number $\nu$, which is an
integer \cite{c1d1,c1d2,c1d3}. The winding number of the Hamiltonian Eq.(\ref{eq2}) is $\nu=1$. In the case, it is equivalent to the Berry phase, which describes the electric polarization. The Berry phase in the $k$ space is defined as: $\gamma=\oint\mathcal{A}(k)dk$
with the Berry connection $\mathcal{A}(k)=i\langle u_{k}|\frac{d}{dk}|u_{k}\rangle$
and $|u_{k}\rangle$ the occupied Bloch states \cite{berry1,berry2}.
Due to the protection of the symmetries in BDI class the Berry
phase $\gamma$ mod $2\pi$ can have two values: $\pi$ for a topologically
nontrivial phase and $0$ for a topologically trivial phase. The topological
property is manifested by the boundary states of zero energy on an open chain.
The spin-down subsystem, which is the time-reversal counterpart of Eq.(\ref{eq2}), has a winding number $\nu=-1$. Then the combined system is time-reversal invariant with the time-reversal operator (the time-reversal operator ${\cal T}$ is the same as the one for Eq.(\ref{eq1}) ).  Then the combined system belongs to DIII class and its topological invariant is a $Z_2$. The $Z_2$ topological invariant for $s_z$ conserved system can be calculated using the berry phase of either spin subsystem.

\section{The oscillatory crossover from $2D$ to $1D$ topological phases}

\begin{figure}[htbp]
\centering \includegraphics[width=7cm]{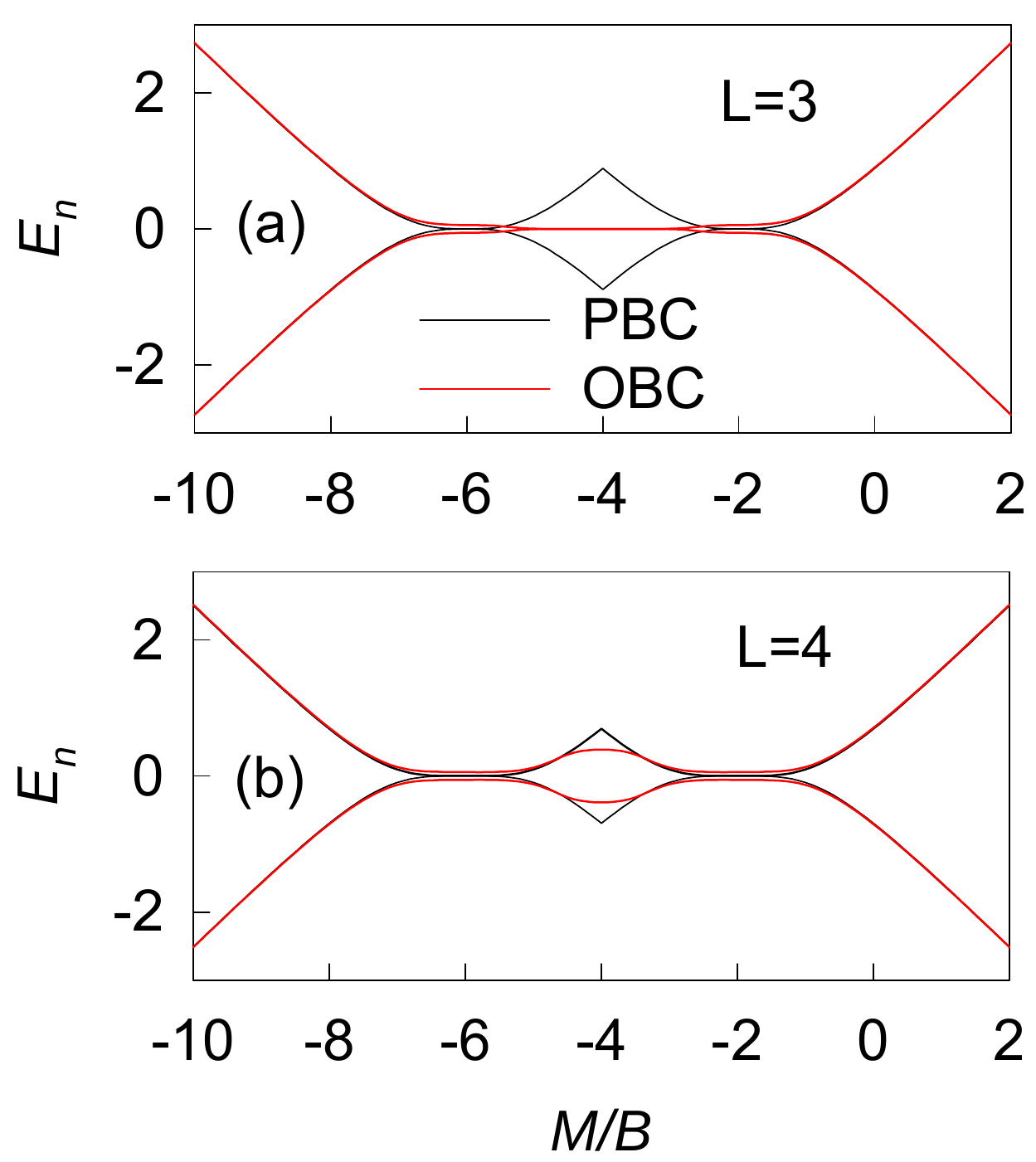} \caption{(Color online) The eigenenergies at half filling and the one above
of Eq.(\ref{eq1}) vs. $M$ on a thin strip with the width: (a) $L=3$;
(b) $L=4$. The length of the strip is $N=50$ with periodic boundary
condition (PBC) or open boundary condition (OBC). }

\label{fig2}
\end{figure}

We consider the 2D TI model Eq.(\ref{eq1}) in a narrow strip configuration.
Its finite-size effect has been studied previously \cite{zhou}. It
is found that on a narrow strip the edge states on the two sides can
couple together to produce a gap in the spectrum. The finite-size
gap $\Delta\propto e^{-\lambda L}$, decays in an exponential law
with the width $L$. The decaying length scale is determined by the
bulk gap of Eq.(\ref{eq1}). As shown in Fig.\ref{fig2}, the
bulk gap vanishes at $M/B=0,-4,-8$, near which the finite-size gap
is maximum. Interestingly the narrow strip as a quasi- $1D$ system
shows $1D$ topological phase in an oscillatory way. It is a $1D$
topological phase with boundary states when the width $L$ is odd, while
trivial when the width $L$ is even. Since spin $s_{z}$ is conserved
in Eq.(\ref{eq1}), the oscillatory crossover happens also for each
spin subsystem. If the spins are coupled (such as by the Rashba spin-orbit
coupling described in Sec. V), the above results still persist.

The oscillatory behavior happens near $M/B=-4$, where the bulk gap
is zero and separates two TI phases. However it is absent near $M/B=0,-8$,
which separate a TI from a trivial insulator. It is noted in Fig.\ref{fig1}
that $M/B=-4$ is deep in the $1D$ topological phase of Eq.(\ref{eq2}),
which is reduced from Eq.(\ref{eq1}). So the oscillatory behavior
is closely related to the corresponding $1D$ topological phase. In
the following section, we construct the $2D$ TI model Eq.(\ref{eq1})
from the point of view of coupled $1D$ models Eq.(\ref{eq2}) to
understand the oscillatory behavior.

\section{Constructing a $2D$ model from $1D$ topological model}

The $2D$ model in Eq.(\ref{eq1}) can be viewed as a set of the coupled
$1D$ models in Eq.(\ref{eq2}). In the limit of zero coupling, the
boundary states of the $1D$ topological model form two flat bands at the
two edges of the $2D$ system, which are topological protected by
the symmetry of the isolated $1D$ model. Next we study the evolution
of the boundary states as the $1D$ chains are coupled by the hopping terms.
Suppose the Hamiltonian of a $1D$ open chain along $x-$ direction
is $H_{1D}^{open,n}$ {[}the same as the one in Eq.(\ref{eq2}){]}
with $n$ denoting the $n-th$ chain (it is identical for different
$n$). Since the Hamiltonian $H_{1D}^{open,n}$ has the chiral
symmetry, its eigenenergies are symmetric to $0$ and we label them: $\{E\}_{n}=-E_{0}^{(n)},E_{0}^{(n)},-E_{1}^{(n)},E_{1}^{(n)},...(E_{0}^{(n)}<E_{1}^{(n)}...)$,
which correspond to the eigenstates: $\{\varphi\}_{n}=\varphi_{0}^{n,-},\varphi_{0}^{n,+},\varphi_{1}^{n,-},\varphi_{1}^{n,+},...$
. For the $2D$ system containing $N_{y}$ $1D$ chains, we choose
the basis: $\Phi=(\{\varphi\}_{0},\{\varphi\}_{1},...,\{\varphi\}_{N_{y}})$.
Under this basis, the $2D$ Hamiltonian can be calculated and has
the following structure,
\begin{eqnarray*}
H_{2D}^{(N_{y})}=\left(\begin{array}{cccc}
H_{diag} & H_{ndiag} & 0 & ...\\
H_{ndiag}^{\dagger} & H_{diag} & H_{ndiag} & ...\\
0 & H_{ndiag}^{\dagger} & H_{diag} & ...\\
0 & 0 & H_{diag} & ...
\end{array}\right),
\end{eqnarray*}
where $H_{diag}=\langle\{\varphi\}_{n}|H_{1D}^{open,n}|\{\varphi\}_{n}\rangle$
is the diagonal matrix with the diagonal elements $\{E\}_{n}$; $H_{ndiag}=\langle\{\varphi\}_{n}|H_{couple}|\{\varphi\}_{n+1}\rangle$
with $H_{couple}$ which describes the coupling  between nearest-neighbor
(NN) chains. $H_{ndiag}$ contains the hopping amplitudes of the eigenstates
between the NN chains, and generally one eigenstate couples with all
other eigenstates of the NN chain.

We firstly consider a special case $M=-4B$, when the zero modes ($E_{0}^{(n)}=0$)
of the $1D$ open chain distribute only on the end sites. Each chain
has two zero modes, one of which is on one end and the other is on
the other end. When only the $A$-kind hoppings with the amplitude $A$ couple
the chains, the zero modes only couple the zero modes on the same
end of the NN chain and the amplitude is $\pm IA$. So the low-energy
physics is described by the zero modes which hop on the edge with
the amplitude $\pm IA$.

For the case of two chains, we have two zero modes on each side. We
can limit to a subspace composed of the zero modes, i.e., ($\varphi_{0}^{1,-},\varphi_{0}^{2,-}$).
Under this basis, the effective Hamiltonian becomes a $2\times2$
matrix:
\begin{eqnarray*}
H_{eff}^{(2)}=\left(\begin{array}{cccc}
0 & iA\\
-iA & 0
\end{array}\right),
\end{eqnarray*}
Its eigenenergy is $E_{eff}^{(2)}=\pm A$ and the system is gapped.
Similarly for the case of three chains, under the basis $(\varphi_{0}^{1,-},\varphi_{0}^{2,-},\varphi_{0}^{3,-})^{T}$,
the effective Hamiltonian becomes a $3\times3$ matrix:
\begin{eqnarray*}
H_{eff}^{(3)}=\left(\begin{array}{cccc}
0 & iA & 0\\
-iA & 0 & iA\\
0 & -iA & 0
\end{array}\right).
\end{eqnarray*}
This matrix has an eigenenergy $0$ with the engenvector $\psi_{0}=\frac{1}{\sqrt{2}}(\varphi_{0}^{1,-}+\varphi_{0}^{2,-})$.
For the case of multi- chains, the resulting effective matrix has
similar structure, i.e., a tridiagonal matrix with zero diagonal elements. The eigenenergy of such Hermitian matrix is symmetric to $0$. So if the dimension of the matrix is odd, it must
have the eigenvalue $0$. The above result can be understood qualitatively: since two coupled boundary states tend to destroy each other,
one can survive from odd number of boundary states.

The above result persists when finite $B$-kind hopping is included.
The $B$-kind hopping couple the zero modes with a few other modes.
For relatively small $B$, the coupling among the zero modes still
dominates. A proper unitary transformation can move the few coupled
modes to one side of the matrix, and the zero modes to the other side.
If the length of the chain is long enough, i.e., the dimension of
the matrix is big enough, their effect on the zero modes can be neglected,
which is known as the finite-size effect. So the oscillating behavior
of the topological property of quasi-$1D$ strip created by narrowing
a $2D$ TI can be understood in the above way.

\begin{figure}[htbp]
\centering \includegraphics[width=8cm]{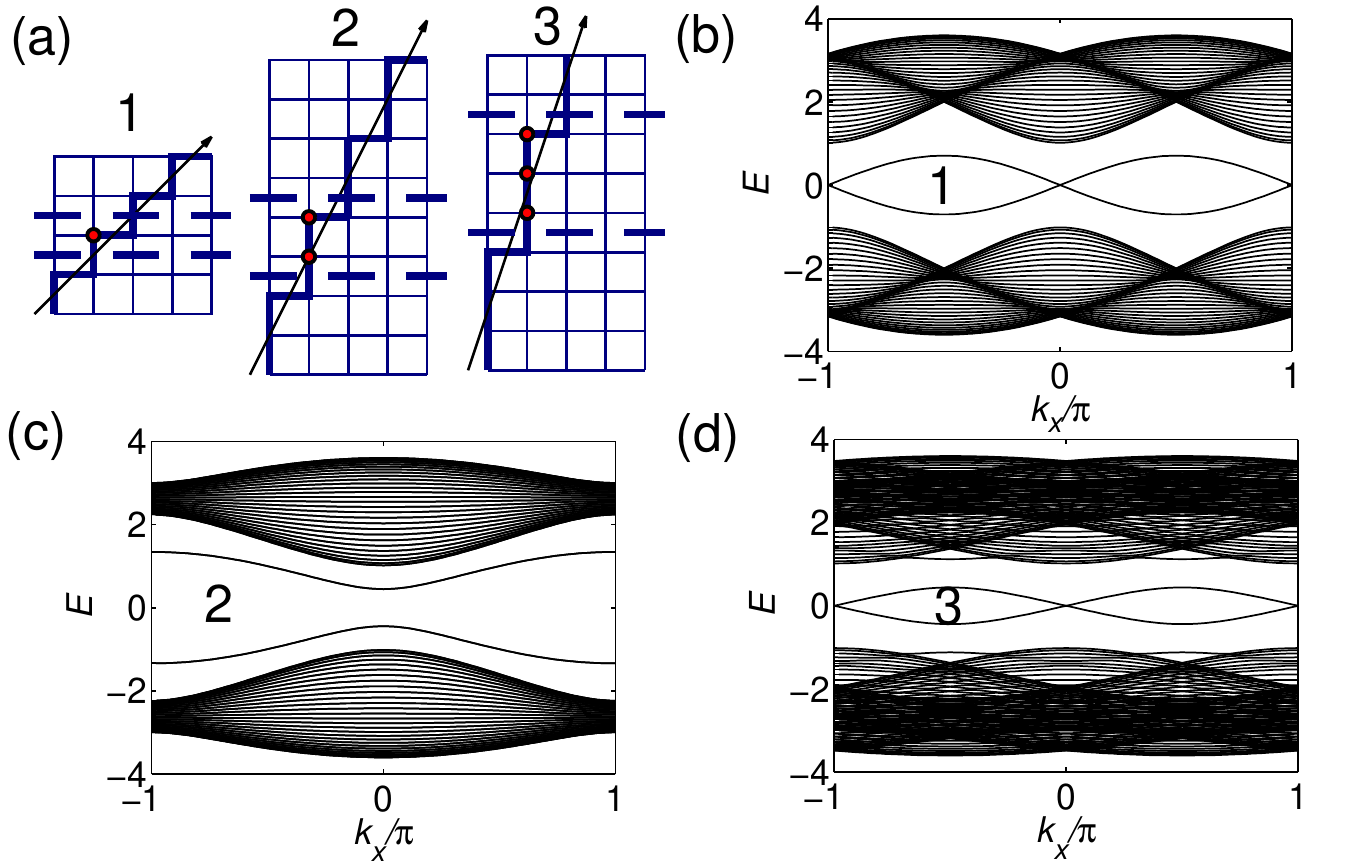} \caption{(Color online) One-dimensional energy bands for a strip with different
edges. The parameters are: $M=-3B$, $B'=0$, when the system is in
the WTI 1 phase.}
\label{fig3}
\end{figure}

In a general case of $N_{y}$, the zero modes form a $1D$ conducting
chain and its energy spectrum is: $E=2A\sin(k_{y})$ with $k_{y}=2\pi/N_{y}$
(the open boundary condition along $x-$ direction and the periodic boundary condition along $y-$ direction). The edge state is a mid-gap one.
 Specially the above edge states
appear on the edge along the $y-$ direction, but disappear on the
edge along $x-$ direction.  For other kinds
of edges, they can be understood from the view of coupling thin strips.
As an example, we consider there different edges shown in Fig.\ref{fig3}
(a). There are mid-gap edge states on the edges $1$ and $3$, but
there are not on the edge $2$. The thin strip in the $x-$ direction
is shown by the dashed lines in the figure. For the system with the
edge $2$, the thin strip contains two chains. As discussed in the
previous section, the boundary modes are gapped, thus there are no mid-gap edge
states when the thin strips are coupled along the $y-$ direction.
While for the system with the edges $1$ or $3$, the thin strip contains
odd number of chains. So the boundary modes persist and there are mid-gap
edge states.

Since the bulk system is gapped, the appearance of the mid-gap edge
states on specific boundaries is due to the topological property
of the bulk insulator. However it is in contrast to the topological property
of quantum spin Hall insulators, where there are always gapless edge states traversing
the gap. The topological invariant of the
above bulk system is $Z_{2}=0$, which is trivial. However the above
phase is different from a trivial insulator. This implies that for
$2D$ time-reversal invariant insulators besides the trivial insulators
and the quantum spin Hall insulators, there exist another insulators,
in which $Z_{2}=0$, but the mid-gap edge states appear on specific boundaries.
It is somehow similar to the $3D$ weak TI phase, which can be understood as layered quantum spin Hall effect. So we term the above phase as '$2D$ weak TI'.

\section{The $2D$ weak TI}

\begin{figure}[htbp]
\centering \includegraphics[width=7cm]{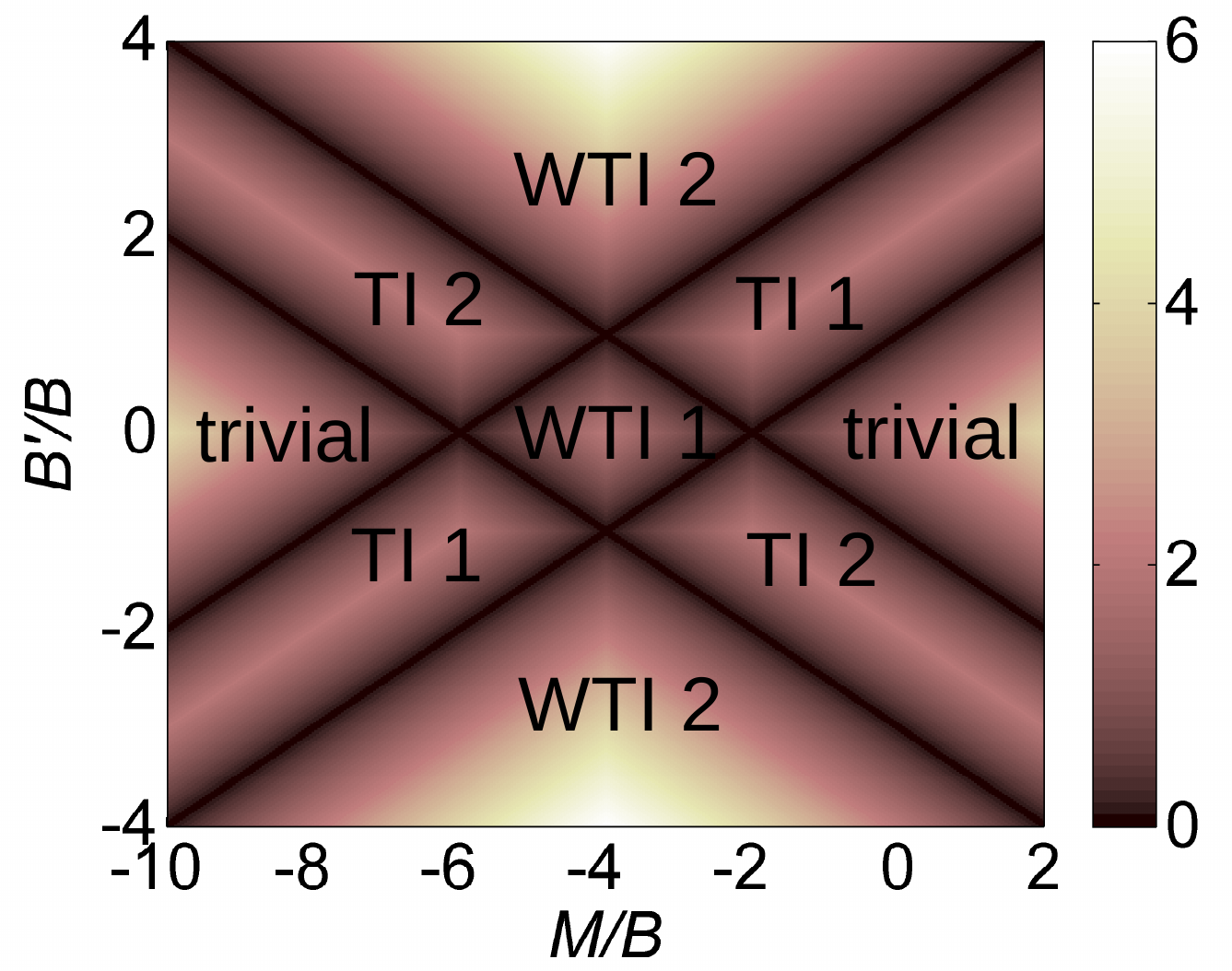} \caption{The phase diagram of the modified Hamiltonian Eq.(\ref{eq1}) with anisotropic $B$-kind hoppings in the
$(M,B')$ plane. In TI 1 (TI 2) the crossing of the edge states is at $k_y=0 (\pi)$ with $y-$ directed boundary. The color represents the gap of the bulk system.}
\label{fig4}
\end{figure}

\begin{figure}[htbp]
\centering \includegraphics[width=7cm]{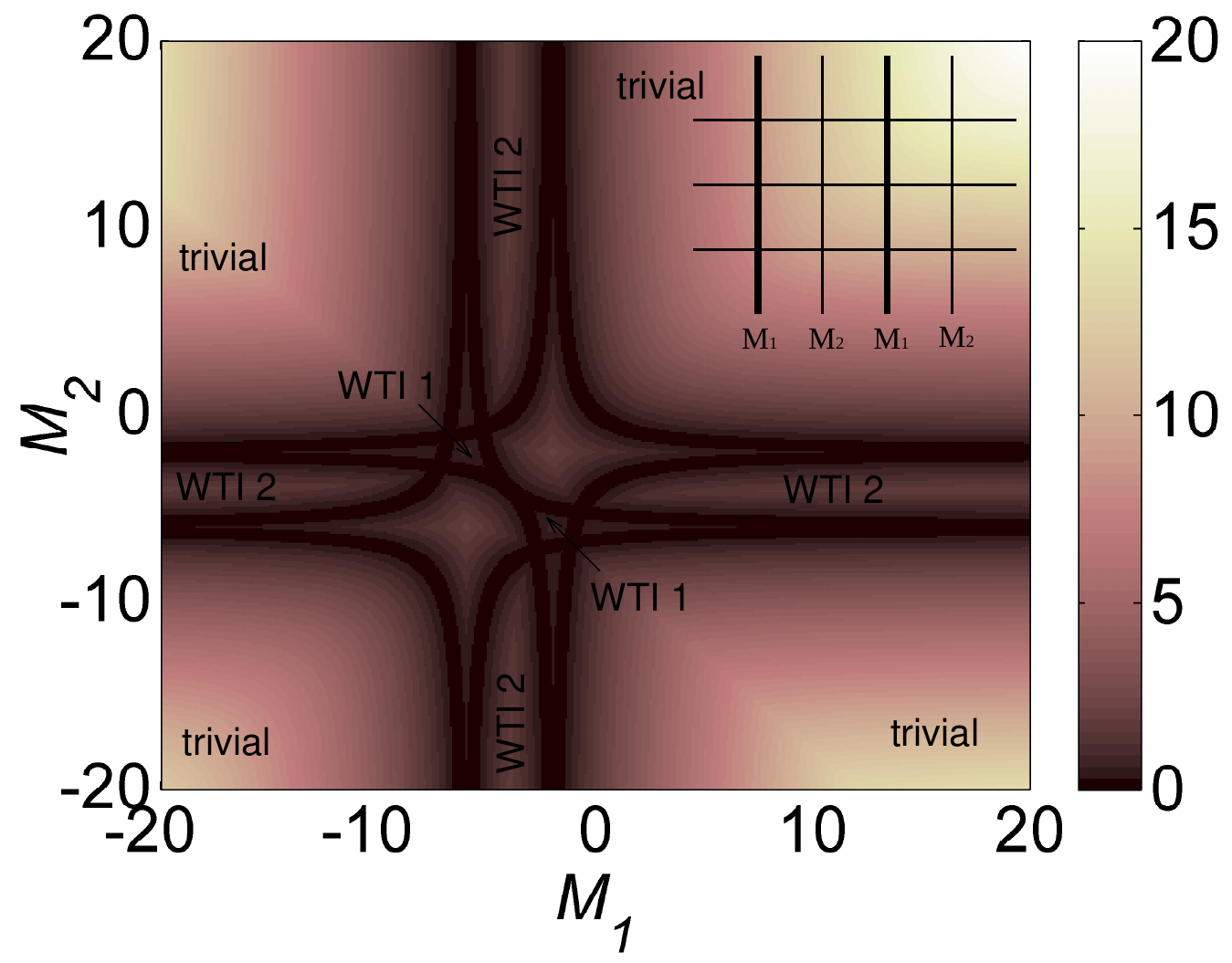} \caption{The phase diagram of the modified Hamiltonian Eq.(\ref{eq1}) with anisotropic mass in the $(M_1,M_2)$ plane. The inset schematically shows the pattern of the mass on the lattice. The regions without notation represent the quantum spin Hall insulator. The color represents the gap of the bulk system.}
\label{fig5}
\end{figure}

In the previous section a weak $2D$ TI is identified with the mid-gap edge
states along specific edges and the topological invariant $Z_{2}=0$. More generally the Hamiltonian in Eq.(\ref{eq1}) can be modified
by changing the amplitude of the $B$-kind hopping along $y-$ direction
to $B'$ which can be tuned. With the points on which the gap closes, the phase diagram of
the modified Hamiltonian in the $(M,B')$ plane can be obtained. As
shown in Fig.\ref{fig4}, besides the quantum spin Hall insulators and trivial insulators,
the $2D$ weak TI exists in three regions of the phase diagram. We distinguishes
the $2D$ weak TIs with the mid-gap edge states appearing on $x-$ edge (WTI 2)
or $y-$ edge (WTI 1), and the quantum spin Hall insulators with the
gapless crossing appearing at $k_y=0$ (TI 1) or $k_y=\pi$ (TI 2).

It is noticed that the $2D$ weak TI exists in the regions with anisotropic $B$-kind hoppings ($B' \neq B$). Indeed the anisotropy is key to realize the phase \cite{hat}. The anisotropy can also be induced in the parameter $M$ of Eq.(\ref{eq1}). Consider the case shown in Fig.\ref{fig5}: the mass $M$ is uniform along $y-$ direction, but has alternating values $M_1,M_2$ along $x-$ direction. The phase diagram in the $(M_1,M_2)$ plane is shown in Fig.\ref{fig5}, in which the $2D$ weak TI is identified in six regions.

Till now the $2D$ weak TI is identified in the anisotropic systems. Its topological invariant $Z_2$ is zero, but the phase is different from the $Z_2=0$ trivial insulators. It is desirable to characterize its topological property. It has been known that the $Z_{2}$ topological invariant of a 2D time-reversal
invariant insulator is defined as: $(-1)^{\nu_{0}}=\prod_{n_{j}=0,1}\delta_{n_{1}n_{2}}$,
with $\delta_{n_{1}n_{2}}$ the time-reversal polarization at the
four time-reversal invariant momenta $\Gamma_{i=(n_{1}n_{2})}=(n_{1}\pi\hat{x}+n_{2}\pi\hat{y})$,
with $n_{j}=0,1$. The above constructed $2D$ Hamiltonian has the
inversion symmetry $H(-{\bf k})=\hat{P}H({\bf k})\hat{P}$ with $\hat{P}$ the inversion operator. In the presence of the inversion symmetry,
$\delta_{i}$ can be determined by the parity of the occupied band
eigenstates: $\delta_{i}=\prod_{m=1}^{N}\xi_{2m}(\Gamma_{i})$, where
$\xi_{2m}$ is the parity eigenvalues of the $2m$-th occupied states.
 The topological property of $2D$ time-reversal invariant insulators is determined by four time-reversal polarizations $\delta_i$. In general $\delta_i$ is not gauge invariant, while in the presence of inversion symmetry $\delta_i$ is gauge invariant. So for the $Z_2=0$ time-reversal invariant insulators with inversion symmetry, the details of the four $\delta_i$ should distinguish the $2D$ weak TIs and the trivial insulators.

\begin{figure}[htbp]
\centering \includegraphics[width=7cm]{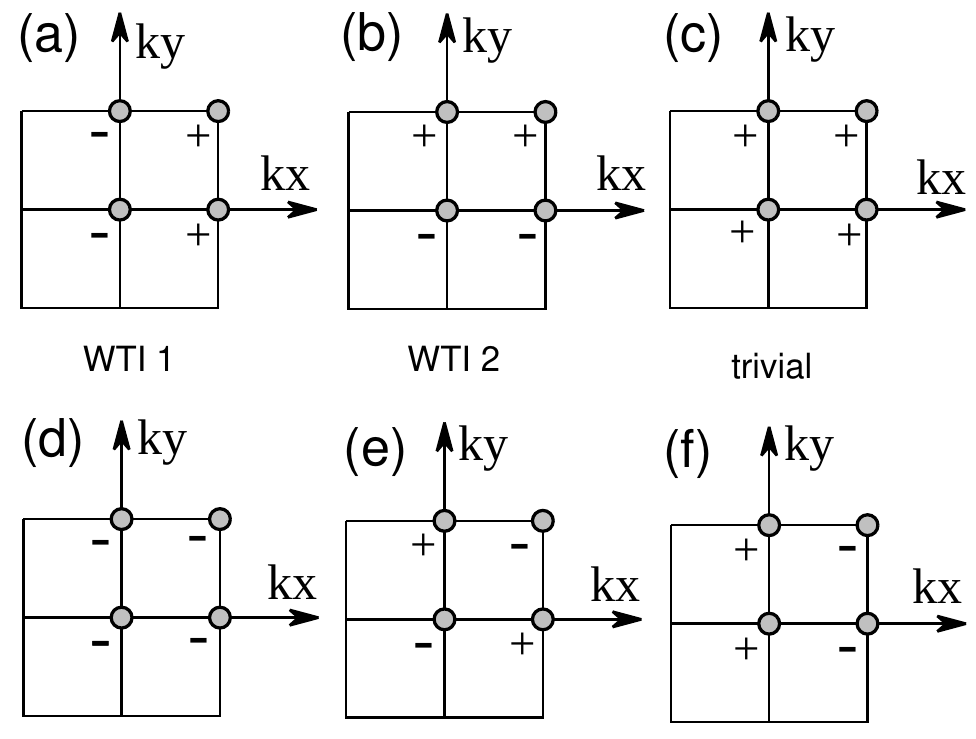} \caption{Depicts $\delta_i$ at the time-reversal invariant momenta of $2D$ weak TIs and trivial insulators. The upper ones are for the case of anisotropic $B$-kind hoppings: (a) WTI 1;(b) WTI 2;(c) trivial insulator. The lower ones are for the case of anisotropic mass: (d) WTI 1;(e) WTI 2;(f) trivial insulator.}
\label{fig6}
\end{figure}

For the case of anisotropic $B$-kind hoppings, the inversion operator is $\hat{P}=I\otimes\sigma_{Z}$ and  $\delta_{i}=-sgn(M(\Gamma_{i}))$ with $\tilde{M}({\bf k})=M+4B-2B\cos(k_{x})-2B'\cos(k_{y})$. In WTI 1, if the boundary is along $y-$ direction, the two $\delta_i$ projected on $k_y=0 (\pi)$ have different signs, which is in contrast to the trivial insulator. Define the $Z_2$ invariant $\pi_{k_{\mu}}$ the product of two $\delta_{i}$ on the line $k_{\mu}$ ($k_{\mu}=n_{i}\pi,\mu=x,y$), then the additional $Z_2$ indices $\pi_{k_{\mu}}$ distinguish the $2D$ weak TI and trivial insulators. $\pi_{k_{\mu}}$ is directly related to the existence of the edge states on $\mu$-directed boundary and $\pi_{k_{\mu}}=-1$ means a crossing of the edge states at $k_{\mu}$. For example in the WTI 1 phase shown in Fig.\ref{fig6}(a), if the boundary is along $y-$ direction, $k_y$ remains good quantum number and $\pi_{k_{y}=0}=\pi_{k_{y}=\pi}=-1$. So there appear mid-gap edge states with two crossings at $k_y=0,\pi$ on the boundaries. Also in quantum spin Hall insulators, $\pi_{k_{\mu}}$ determines the position of the crossing of the edge states, which happens at $k_{\mu}$ with $\pi_{k_{\mu}}=-1$.

For the case of anisotropic mass, the inversion operator is $\hat{P}=I\otimes diag(\sigma_z, e^{-i k_y/2}\sigma_z)$ (the inversion center is chosen on a site with the mass $M_1$). The $\delta_i$ of the $2D$ weak TI phases are calculated, which is shown in Fig.6. It seems that the above discussion is inapplicable. Actually to characterize the topological property properly, the $\delta_i$ should be defined compared to the corresponding $\delta_i^{trivial}$ of the trivial insulator in the same model, i.e., $\tilde{\delta_i}=\delta_i \delta_i^{trivial}$. With $\tilde{\delta_i}$, Fig.\ref{fig6}(d) [(e)] is the same as Fig.\ref{fig6}(a) [(b)], and the $2D$ weak TI is characterized correctly.

\begin{figure}[htbp]
\centering \includegraphics[width=8.5cm]{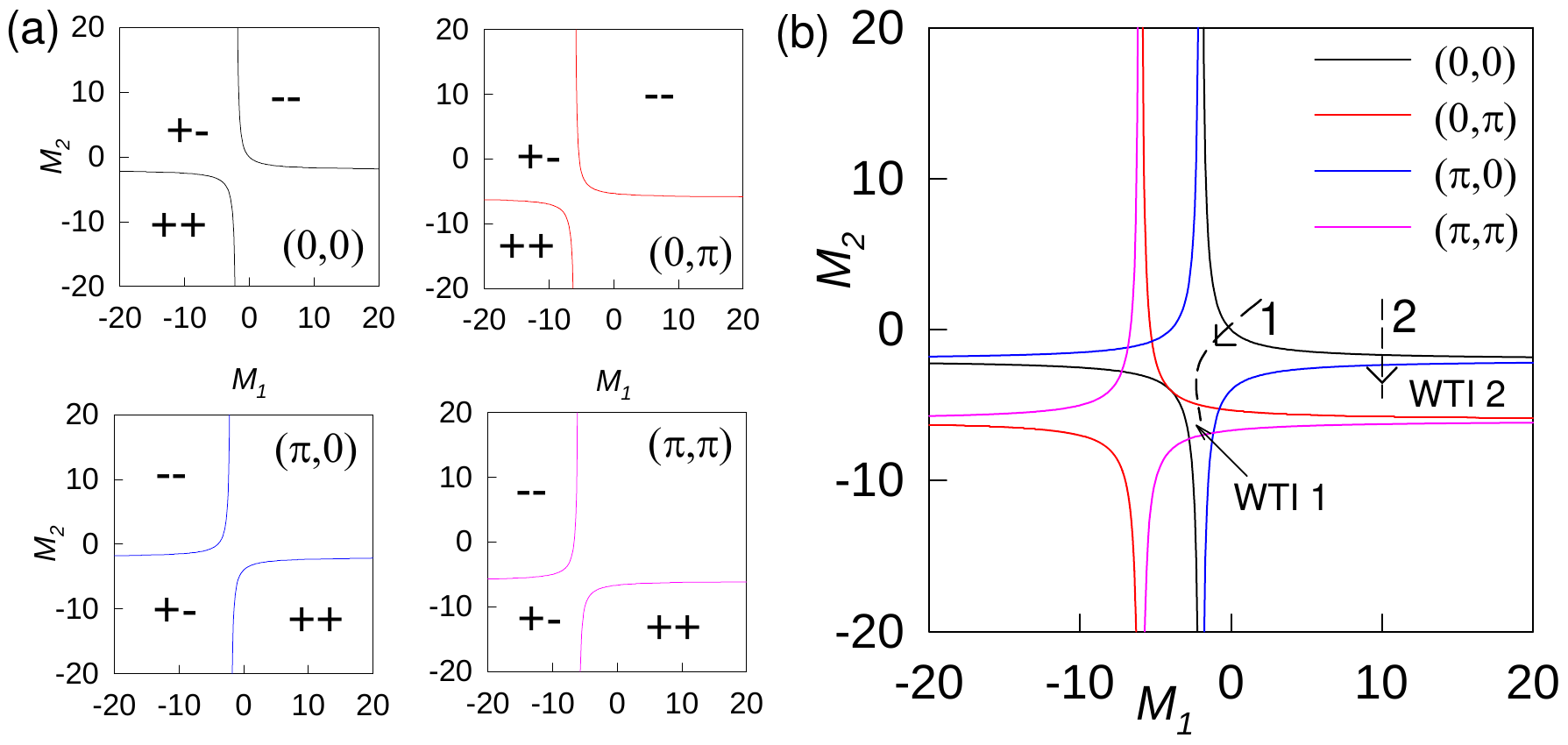} \caption{(Color online) (a) The zero-gap line in $(M_1,M_2)$ plane for the case of anisotropic mass at the four time-reversal invariant momenta. $+/-$ is the parity of the occupied band. (b) The phase diagram obtained by combining the zero-gap lines in (a). The arrow dashed line $1(2)$ is the path to WTI $1(2)$ (the ones corresponding to (d) and (e) in Fig.\ref{fig6}) from a trivial insulator.}
\label{fig7}
\end{figure}

So to correctly characterize the $2D$ weak TI with inversion symmetry, the $\delta_i$ should be defined compared to that of the trivial insulator in the same model. It can be understood from the view of band inverting. We take the case of anisotropic mass as an example. Its Hamiltonian in the momentum space writes as,
\begin{eqnarray}
H_{M}({\bf k})=\left(\begin{array}{cccc}
h_1(k_y) & h_{12}(k_x)\\
h_{12}^{\dagger}(k_x) & h_2(k_y)
\end{array}\right),
\end{eqnarray}
where $h_{1,2}(k_y)=[M_{1,2}+4B-2B\cos(k_y)]\sigma_z+2A\sin(k_y)\sigma_x$ and $h_{12}(k_x)=-B(1+e^{-ik_x})\sigma_z-iA(1-e^{-ik_x})\sigma_y$. The eigenenergies and eigenvectors at the time-reversal invariant momenta can be obtained analytically. In Fig.\ref{fig7} (a) the zero-gap line in the $(M_1,M_2)$ plane at each time-reversal invariant momentum and the parities of the occupied bands are shown. When the zero-gap line is crossed, there occurs a band inverting. By combining all the zero-gap lines, the phase diagram in Fig.\ref{fig5} is recovered. Any phase in the phase diagram can be reached by band inverting starting from a trivial insulator. Thus $\tilde{\delta_i}=\delta_i \delta_i^{trivial}$ records the number of band inverting from a trivial insulator. $\tilde{\delta_i}=-1$ means an odd number of band inverting and there appears a crossing of the edge state at $i$-th time-reversal invariant momentum. Then the topological property can be analyzed correctly with $\tilde{\delta_i}$.

For the general case without inversion symmetry, the topological property of the $2D$ weak TI can be understood from the Berry phase of one spin subsystem, since the $2D$ weak TI is closely related to $1D$ topological phase. The Berry phase defined with $k_y(k_x)$ at fixed $k_x(k_y)$ can be calculated. The symmetries which protects the $1D$ topological phase is broken except at specific $k_x(k_y)$. If there are two $k_x(k_y)$ at which the Berry phase is $\pi$, which means the edge states exist and have two crossings in the presence of $x(y)$-directed boundary, the system is a $2D$ weak TI.

The above analysis is based on the system with spin $s_{z}$ conservation.
However the result is applicable to any time-reversal invariant insulators.
Next we study the case of the spin-up and -down subsystem coupled
by Rashba spin-orbit coupling, which preserve the time-reversal invariant
symmetry,
\begin{eqnarray}
H_{R}=&-&\sum_{i,\hat{x}}sgn(\hat{x})i\lambda_{R}\Psi_{i}^{\dagger}s_{y}\otimes I\Psi_{i+\hat{x}}\label{eq5}\\
 & + & \sum_{i,\hat{y}}sgn(\hat{y})i\lambda_{R}\Psi_{i}^{\dagger}s_{x}\otimes I\Psi_{i+\hat{y}}.\nonumber
\end{eqnarray}
Adding the above term to the modified version of the Hamiltonian Eq.(\ref{eq1}).
For the case of anisotropic $B$-kind hoppings,
the total Hamiltonian in the momentum space is,
\begin{eqnarray}\label{eq6}
H_{T}({\bf k})&=&\tilde{M}({\bf k})I\otimes\sigma_{z}+\tilde{A}(k_{x})s_{z}\otimes\sigma_{x}+\tilde{A}(k_{y})I\otimes\sigma_{y}\nonumber \\
&+&\tilde{\lambda_{R}}(k_{x})s_{y}\otimes I-\tilde{\lambda_{R}}(k_{y})s_{x}\otimes I.
\end{eqnarray}
Its energy spectrum is:
\begin{eqnarray}
(E_{T}({\bf k}))^{2}=[\tilde{A}(k_{x})]^{2}+\{\pm\sqrt{[\tilde{\lambda_{R}}(k_{x})]^{2}+[\tilde{\lambda_{R}}(k_{y})]^{2}}\label{eq7}\\
+\sqrt{[\tilde{M}({\bf k})]^{2}+[\tilde{A}(k_{y})]^{2}}\}^{2}.\nonumber
\end{eqnarray}
with $\tilde{A}(k_{x})=2Asin(k_{x})$, $\tilde{A}(k_{y})=2Asin(k_{y})$,
$\tilde{\lambda_{R}}(k_{x})=2\lambda_{R}\sin(k_{x})$, $\tilde{\lambda_{R}}(k_{y})=2\lambda_{R}\sin(k_{y})$.
As has been known,  the Rashba spin-orbit coupling does not break
the quantum spin Hall effect when it is small. In the following we
show that the weak TI is also robust to it. From Eq.(\ref{eq7}),
it is noticed that the gap closing is independent of $\lambda_{R}$
for $\lambda_{R}<A$. Since the topological quantum phase transition
occurs when the gap closes, the weak TI is not affected by the Rashba
spin-orbit coupling in this case. For $\lambda_{R}>A$, the gap of
the bulk system vanishes at a critical $\lambda_{R}^{c}$, when the
weak TI is broken. The calculated energy spectrum and time-reversal polarization $\delta_i$ are consistent with the above analysis. For the case of anisotropic mass, the result is similar.

\begin{figure}[htbp]
\centering \includegraphics[width=8.5cm]{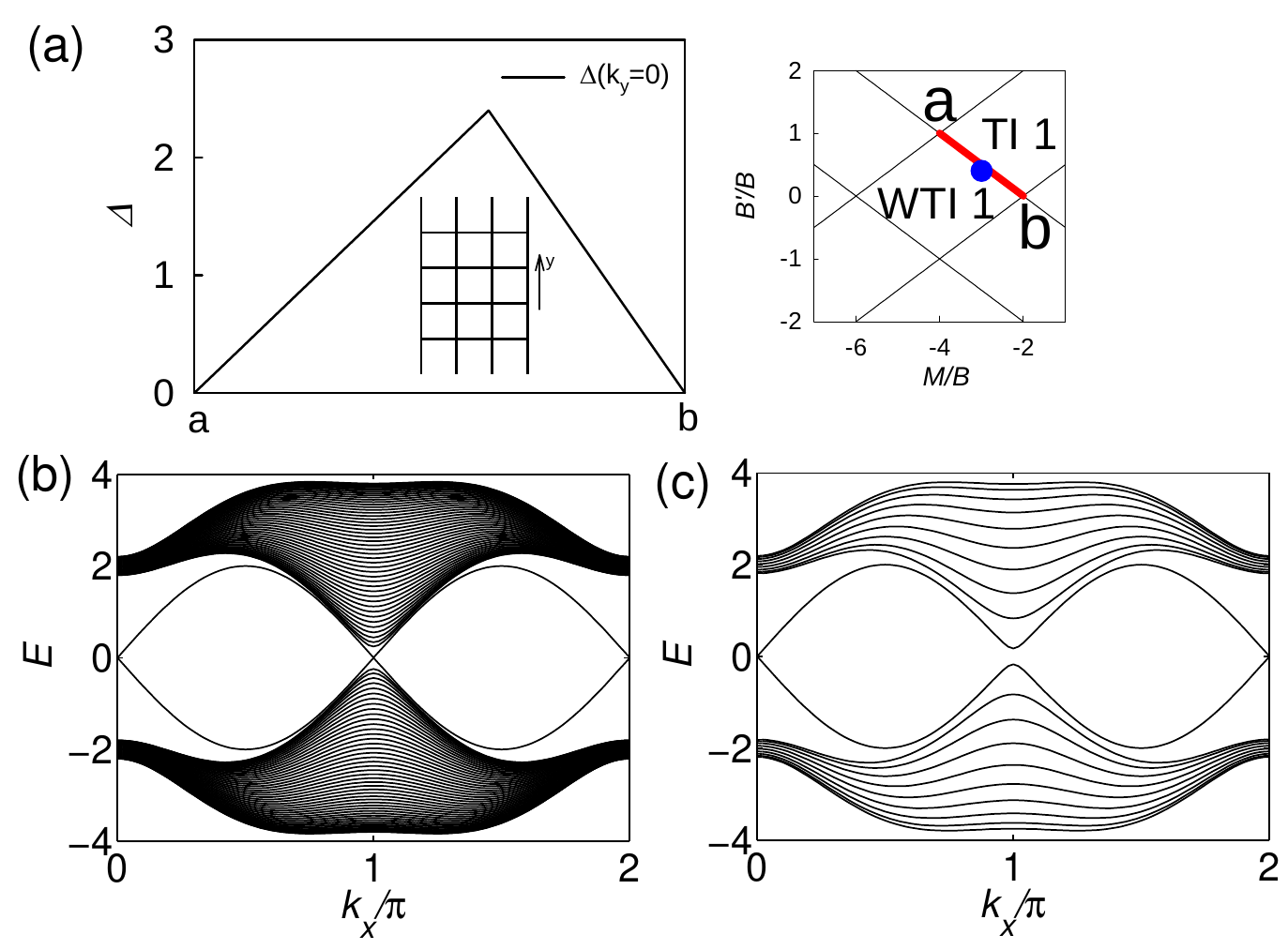} \caption{(Color online) (a) The bulk gap at $k_{y}=0$ with the parameters along the boundary
of WTI 1 and TI 1 in the phase diagram (the bulk gap at $k_{y}=\pi$
is all zero). The energy spectrum on a $y$-directed strip with the width: (b) $L=50$
and (c) $L=10$. The parameters are: $M/B=-3,B'/B=0.4$, which is
denoted by a blue dot in the right figure of (a) and it is in the WTI 1 phase
near the boundary.}

\label{fig8}
\end{figure}

As has been stated, since the mid-gap edge states in the $2D$ weak TI are related to the corresponding $1D$ boundary modes, it can be destroyed
by the disorder. However under specific conditions, they
can be robust as those in the $Z_{2}=1$ quantum spin Hall insulators.
It can happen in the region near the boundary between WTI and TI phases. For example,
for the case with anisotropic $B$-kind hoppings, in the middle of the boundary between WTI and TI phases, the gaps at the two valleys $k_{x (y)}=0,\pi$ is different. For the case shown
in Fig.\ref{fig8}, the gap at $k_{y}=\pi$ is nearly zero while
is still large at $k_{y}=0$. Considering a narrow strip with its edges
along $y-$ direction, due to the finite-size effect, the edge
states at $k_{y}=\pi$ is gapped, while the ones at $k_{y}=0$ persist. Thus a single robust gapless crossing
at $k_{y}=0$ is realized in the finite-size gap.

\section{Conclusions and Discussions}

Dimensional evolution between $1D$ and $2D$ topological phases is
investigated systematically. The crossover from a $2D$ TI to its
$1D$ limit shows oscillatory behavior between a $1D$ ordinary insulator
and $1D$ TI. By constructing a $2D$ topological system from $1D$
TI, it is shown that there exist the weak topological phase in $2D$ time-reversal
invariant band insulators. The phase can be realized in anisotropic systems. In the weak phase, the topological invariant
$Z_{2}=0$ and the edge states only appear along specific boundaries.
Since the edge states are closely related to the boundary states of the
corresponding $1D$ topological phase, they may be destroyed by disorder
and have symmetry-protecting nature as the corresponding $1D$ topological
phase. The effect of the Rashba spin-orbit coupling,
which preserves time-reversal invariant symmetry, but couples the
spins, is also studied. These results provide further
understanding on $2D$ time-reversal invariant insulators.

Finally we discuss the relevance of the results to experimental measurements.
It is unclear whether the $2D$ weak TI materials exist in nature.
However since the anisotropy is important, it should be searched in
anisotropic materials. Besides real materials, recently the double-well
potential formed by laser light has been developed \cite{exp1}, in
which $s$ and $p$ orbital cold-atoms can be loaded. It has been
shown that a $1D$ topological model similar to Eq.(\ref{eq2}) can
be derived from the experimentally realized double-well lattices by
dimension reduction \cite{1d4}. Another experimental platform is
the photonic quasicrystals, on which the topological properties have
been studied in an engineered way \cite{1d2}. With the fine tuning
of the parameters and the geometries in these experiments, the present
results are very possibly realized experimentally.

\section*{Acknowledgments}

The authors thank Hua Jiang and Juntao Song for helpful discussions. This work was supported by NSFC under Grants No.11274032 and No. 11104189, FOK YING TUNG EDUCATION FOUNDATION, Program for NCET (H.M.G), and the Research
Grant Council of Hong Kong under Grant No. HKU 703713P (S.Q.S).

\textit{Note added.-}
{Upon finalizing the manuscript we noticed a recent preprint \cite{note} on closely related topics.

\end{document}